\documentclass[useAMS,usenatbib]{mn2e}
\usepackage{graphicx}
\usepackage{natbib}
\usepackage[T1]{fontenc}
\usepackage{aecompl}

\def\Mwd{M$_{\mathrm{WD}}$}
\def\Msun{M$_{\odot}$}
\def\Rsun{R$_{\odot}$}
\def\Teff{$T_{\mathrm{ef{}f}}$}

\def\uprime{u$^{\prime}$}
\def\gprime{g$^{\prime}$}
\def\rprime{r$^{\prime}$}
\def\iprime{i$^{\prime}$}
\def\zprime{z$^{\prime}$}

\let\textbf\textrm

\title[Precise parameters for both white dwarfs 
	in the eclipsing binary CSS\,41177]
	{Precise parameters for both white dwarfs 
	in the eclipsing binary CSS\,41177}
\author[M.C.P. Bours et al.]
  {M.C.P.~Bours$^1$\thanks{m.c.p.bours@warwick.ac.uk},
  T.R.~Marsh$^1$,
  S.G.~Parsons$^2$,
  C.M.~Copperwheat$^3$,
  V.S.~Dhillon$^4$,
  \newauthor 
  S.P.~Littlefair$^4$,
  B.T.~G\"ansicke$^1$,
  \textbf{A.~Gianninas}$^5$,
  \textbf{P.-E.~Tremblay}$^6$. \\
  $^1$Department of Physics, Gibbit Hill Road, University of Warwick, Coventry CV4 7AL, UK \\
  $^2$Departmento de F\'isico y Astronom\'ia, Universidad de Valpara\'iso, Avenida Gran Bretana 1111, Valpara\'iso, Chile \\
  $^3$Astrophysics Research Institute, Liverpool John Moores University, Twelve Quays House, Birkenhead, CH41 1LD, UK \\
  $^4$Department of Physics and Astronomy, University of Sheffield, Sheffield, S3 7RH, UK \\
  $^5$\textbf{Homer L. Dodge Department of Physics and Astronomy, University of Oklahoma, 440 W. Brooks St., Norman, OK 73019, USA} \\
  $^6$\textbf{Space Telescope Science Institute, 700 San Martin Drive, Baltimore, MD,
21218, USA} }

\date{Accepted XX XXXX -- Received XX XXXX}

\pagerange{\pageref{firstpage}--\pageref{lastpage}} \pubyear{2002}

\def\LaTeX{L\kern-.36em\raise.3ex\hbox{a}\kern-.15em
    T\kern-.1667em\lower.7ex\hbox{E}\kern-.125emX}

\begin{document}

\label{firstpage}

\maketitle

\begin{abstract}
We present ULTRACAM photometry and X-Shooter spectroscopy of the eclipsing double white dwarf binary CSS\,41177, the only such system that is also a double-lined spectroscopic binary. Combined modelling of the light curves and radial velocities yield masses and radii for both white dwarfs without the need to assume mass-radius relations. We find that the primary white dwarf has a mass of $M_1$ = 0.38 $\pm$ 0.02 \Msun~and a radius of $R_1$ = 0.0222 $\pm$ 0.0004 \Rsun. The secondary white dwarf's mass and radius are $M_2$ = 0.32 $\pm$ 0.01 \Msun~and $R_2$ = 0.0207 $\pm$ 0.0004 \Rsun, and its temperature and surface gravity ($T_2$ = 11678 $\pm$ 313 K, log($g_2$) = 7.32 $\pm$ 0.02) put it close to the white dwarf instability strip. However, we find no evidence for pulsations to roughly 0.5\% relative amplitude. Both masses and radii are consistent with helium white dwarf models with thin hydrogen envelopes of $\leq 10^{-4}$ M$_*$. The two stars will merge in 1.14 $\pm$ 0.07 Gyr due to angular momentum loss via gravitational wave emission.
\end{abstract}


\section{Introduction}
More than 95\% of all stars end their lives as white dwarfs. These most common remnants of stellar evolution trace the initial stellar distribution and contain a history of the evolution of their progenitor stars, through the strong correlation between the progenitor star's initial mass and the white dwarf's final mass \citep{Weidemann00}. As even the oldest white dwarfs have not had enough time to cool below detectability, they provide useful and independent age estimates when found in, for example, the Galactic disc \citep{Wood92}, as well as providing distance estimates to globular clusters \citep{Renzini96}. White dwarfs also feature as the most likely candidates for supernova Type Ia progenitors \citep{Nomoto82, IT84, WI73}, in accreting binaries such as cataclysmic variables, and are potential progenitors of single hot sdB/sdO stars and extreme helium stars \citep{Webbink84}, AM CVn binaries \textbf{\citep{Breedt12}} and supernovae Type .Ia \citep{Bildsten07}.

Despite the abundance and importance of white dwarfs, it has proved difficult to measure fundamental parameters such as mass and radius directly, without the use of theoretical mass-radius relations. This leaves the empirical basis for this relation relatively uncertain \citep{Schmidt96}. For single white dwarfs, spectral fitting can be used to obtain the temperature and surface gravity, after which both mass and radius can be inferred, but only when combined with a mass-radius relation \citep[see for example][]{Provencal02}. White dwarfs in visual binaries, common proper motion pairs, or in open clusters allow one to determine parameters without the use of this relation, therefore providing a direct test of it. These methods rely on accurate parallax measurements, spectral fitting and/or radial velocity measurements \citep{Holberg12, Provencal02, Casewell09}. The number of stars to which these methods can be applied \textbf{is} limited, and with the exception of Sirius B \citep{Barstow05}, they cluster around a mass of \Mwd $\sim$ 0.6 \Msun, making it difficult to test the full range of the mass-radius relation.

Observing white dwarfs in eclipsing binaries enables high precision in determining masses and radii and these types of binaries also include white dwarfs across a wide range of masses \cite[][this paper]{Pyrzas12, OBrien01, Parsons12_419}. For these systems, masses can be determined from orbital velocities and radii from light curve analysis \citep{Parsons12_420, Parsons12_426}. Due to their eclipsing nature, the inclination of the system is strongly constrained, allowing for direct mass determinations as opposed to lower limits. 

The number of known short-period, eclipsing, white dwarf binaries has grown substantially within the last decade, mainly due to large all-sky surveys such as the Sloan Digital Sky Survey \citep[SDSS,][]{York00} and the Catalina Sky Survey \citep[CSS,][]{Drake09}. The latter and other survey work also led to the discovery of the first five eclipsing binaries in which both stars are white dwarfs: NLTT~11748 \citep{Steinfadt10}, CSS\,41177 \citep{Parsons11}, GALEX J171708.5+675712 \citep{Vennes11}, SDSS J065133.33+284423.37 \citep{Brown11} \textbf{and SDSS J075141.18-014120.9 \citep{Kilic13}}. 

These binaries provide us with an opportunity for precise and independent mass and radius measurements for two white dwarfs from one system. The double white dwarf binary that is the subject of this paper, CSS\,41177 (SDSS J100559.10+224932.2), is the only one of the five that is also a double-lined spectroscopic binary, allowing direct measurement of the stars' orbital velocities \citep{Parsons11} and therefore their masses without needing to assume a mass-radius relation.

CSS\,41177 was initially discovered to be an eclipsing binary by \citet{Drake10}, who constrained it to be a white dwarf with a small M-dwarf companion, although they noted that a small faint object could produce a signal similar to what they observed. \citet{Parsons11} obtained Liverpool Telescope + RISE fast photometry and Gemini + GMOS spectroscopy, which allowed them to determine that there were in fact two white dwarfs and to carry out an initial parameter study. In this paper we present higher signal-to-noise data in order to determine the system parameters more precisely and independent of any mass-radius relationships.

\section{Data} \label{sect:data}
\subsection{Photometry: ULTRACAM} \label{subsect:ULTRACAM}
The photometric data were taken with ULTRACAM \citep{Dhillon07}, a visitor instrument that was mounted on the 4.2m William Herschel Telescope (WHT) on the island of La Palma, Spain and on the 3.5m New Technology Telescope (NTT) at the La Silla Observatory, Chile. ULTRACAM is a high-speed camera that images ultraviolet, visual and red wavelengths simultaneously with three frame-transfer CCDs with $\sim$ 25 ms dead time between exposures. Filters identical to the SDSS \uprime, \gprime~and \rprime~filters were used and we windowed the CCD which allowed us to reach exposure times as short as 1.5~seconds in the slow readout speed. We observed eleven primary eclipses of CSS\,41177 and nine secondary eclipses in January 2012 with ULTRACAM mounted on the WHT, including one observation spanning a complete orbit (1.1 cycles). We also observed one primary eclipse in May 2011 with ULTRACAM on the NTT. We refer to a primary eclipse when the hotter, more massive white dwarf is being eclipsed.

The ULTRACAM pipeline \citep{Dhillon07} was used to debias and flatfield the data. The source flux was determined with relative aperture photometry, using a nearby star as a comparison. We used a variable aperture, where the radius of the aperture was scaled according to the full width at half maximum of the stellar profile. The profiles were fit with Moffat profiles \citep{Moffat69}.

The ULTRACAM data has absolute timestamps better than 0.001~seconds. We converted all times onto a TDB timescale, thereby correcting for the Earth's motion around the Solar System barycentre. A code based on SLALIB was used for these corrections, which has been found to be accurate at a level of 50 microseconds compared to TEMPO2 \citep[a pulsar timing package, see][]{Hobbs06}. Compared to the statistical uncertainties of our observations, this is insignificant. 

\subsection{Spectroscopy: X-Shooter}
We obtained spectra with the X-Shooter spectrograph \citep{Vernet11} on the Very Large Telescope (VLT) UT2 ({\it Kueyen}) on the nights of the 25th, 26th and 27th of March 2012, obtaining a total of 117 spectra covering 1.5, 1.4 and 1.7 orbital cycles in the three nights. A log of the observations is given in Table~\ref{tab:obs}. 

The X-shooter spectrograph consists of three independent arms (UVB, VIS and NIR), giving a simultaneous wavelength coverage from 3000 to 24800\AA. We obtained a series of spectra 310, 334 and 300 seconds in length for the UVB, VIS and NIR arms respectively. Spectra were obtained consecutively with occasional short breaks of a few minutes in order to check the position of the target on the slit. We binned by a factor of two both spatially and in the dispersion direction in the UVB and VIS arms, and used slit widths of 0.8\arcsec, 0.9\arcsec and 0.9\arcsec for the UVB, VIS and NIR arms respectively. The NIR arm slit includes a $K$-band blocking filter which reduces the thermal background in the $J$- and $H$-bands.

We reduced these data using version 1.5.0 of the X-shooter pipeline and the Reflex workflow management tool. The standard recipes were used to optimally extract and wavelength calibrate each spectrum. We removed the instrumental response using observations of the spectrophotometric standard star LTT4364 (GJ-440). We obtained and reduced the data in `stare' mode. For optimum sky subtraction in the infrared arm `nod' mode would be preferable, but our priority was to maximise the temporal resolution for the spectral features of interest in the other two arms.

\begin{table}
\caption{Log of the X-shooter observations. On all three nights the conditions were clear, with seeing between 0.5\arcsec and 1.0\arcsec.}
\begin{center}
\label{tab:obs}
\begin{tabular}{l l l l l l}
\hline \hline
Date           &\multicolumn{2}{c}{UT} &\multicolumn{3}{c}{No. of exposures}\\
                &start          &end    &UVB &VIS &NIR \\
\hline \hline
25 March 2012   &00:13          &04:31  &44 &40 &48    \\
26 March 2012   &00:09          &04:10  &39 &35 &42    \\
27 March 2012   &23:51          &04:36  &46 &42 &50    \\
\hline \hline
\end{tabular}
\end{center}
\end{table}

\subsection{Photometry: RISE on the LT}
In order to monitor any long-term orbital period variations we also regularly observed primary eclipses of CSS\,41177 with the fully-robotic 2.0m Liverpool Telescope \citep[LT;][]{Steele04} and RISE camera. RISE contains a frame transfer CCD, with a single `V+R' filter. We observed a total of 10 primary eclipses from February 2011 until March 2013. 

The data were flatfielded and debiased in the automatic pipeline, in which a scaled dark-frame was removed as well. Aperture photometry was performed using the ULTRACAM pipeline in the same manner as outlined in Sect.~\ref{subsect:ULTRACAM}. We converted the resulting mid-eclipse times to the TDB timescale. LT/RISE has absolute timestamps better than 0.1~second (D.~Pollacco, private communication).

\section{Data analysis \& results} \label{sect:analysis&results}
\subsection{Radial velocity amplitudes} \label{sect:spectroscopy}
\begin{figure}
\includegraphics[]{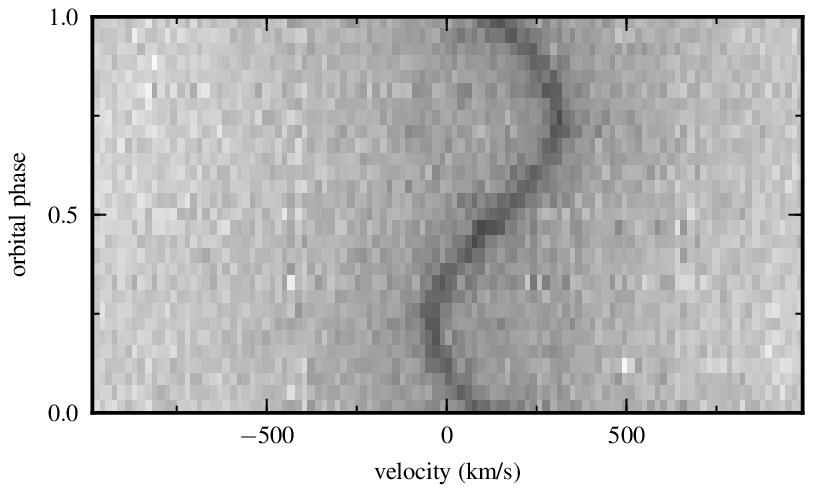}
\includegraphics[]{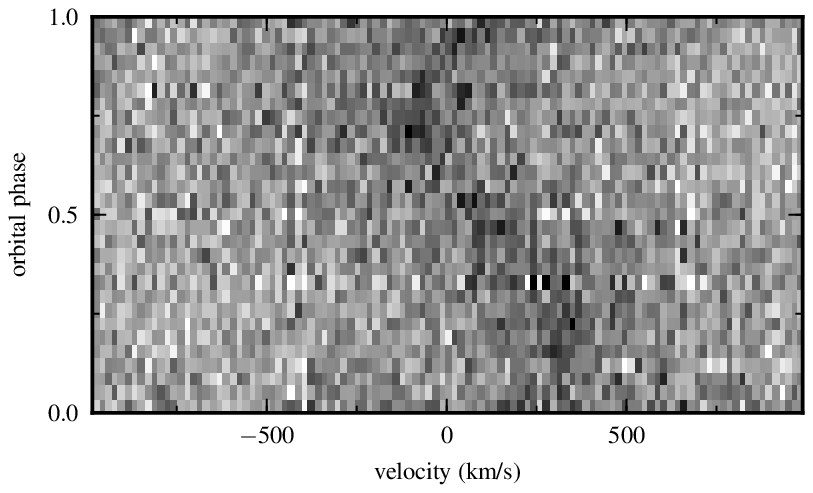}
\includegraphics[]{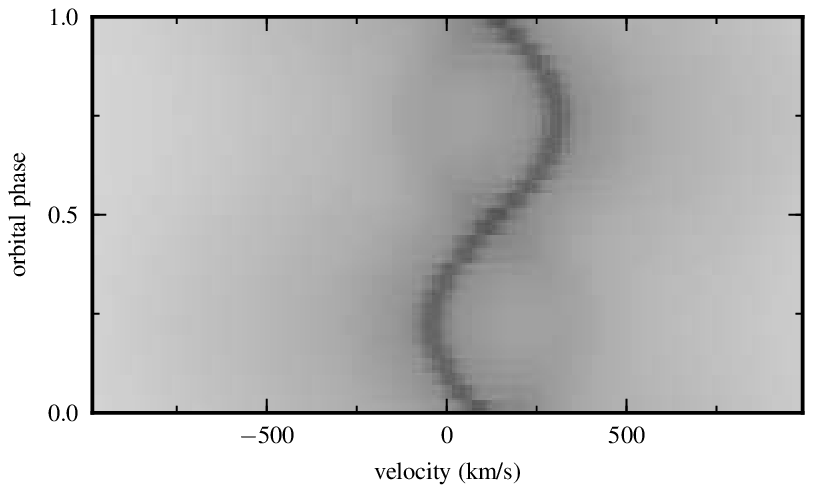} 
\caption{X-Shooter spectroscopy of CSS\,41177. \textit{Top:} spectra of the H$\alpha$ absorption line at 6562.76\AA, folded with the orbital period. The scale is such that black equals 20\% of the continuum and white 100\%. \textit{Middle:} phase-folded spectra after subtraction of the primary white dwarf's H$\alpha$ absorption line, showing the contribution of the secondary white dwarf. Here black equals -18\% of the continuum and white 5\%. \textit{Bottom:} phase-folded model. Black equals 20\% of the continuum and white 100\%.}
\label{fig:X-Shooter_trails}
\end{figure}

We analysed the reduced spectroscopic data using the MOLLY software\footnote{MOLLY was written by T.~R.~Marsh and is available from http://www.warwick.ac.uk/go/trmarsh/software.}, and used observations of the standard star to normalise the continuum and reduce telluric absorption features present in the CSS\,41177 science spectra as far as possible. 

To measure the radial velocity amplitudes $K_1$ and $K_2$ we fitted the H$\alpha$ line with multiple Gaussian profiles combined with a straight line fit to the local continuum using Marquardt's method of minimisation. Each Gaussian profile was represented by two free parameters, the full width at half maximum and the height. The radial velocity $v_r$ was calculated using
\begin{equation}
v_r = \gamma + K_{1,2} \sin(2\pi \phi),
\end{equation}
where $\gamma$ accounts for a systemic radial velocity, $K_1$ and $K_2$ are the radial velocity amplitudes of the two white dwarfs and $\phi$ is the orbital phase. Our best fit gave $K_1 = -176.1 \pm 1.1$ km s$^{-1}$ and $K_2 = 210.4 \pm 6.1$ km s$^{-1}$ for the primary and secondary white dwarfs, and $\gamma = 130.5 \pm 0.7$ km s$^{-1}$ for the offset. The phase-folded trail of the H$\alpha$ line is shown in the top panel in Fig.~\ref{fig:X-Shooter_trails}. From this best model we set the Gaussian profiles corresponding to the secondary star's line to zero, leaving a model of only the primary star's line. We then subtracted the model from the original spectra to bring out the secondary's contribution, shown in the middle panel of Fig.~\ref{fig:X-Shooter_trails}. The systemic radial velocity is clearly visible in both of these figures as the offset from zero. The bottom panel shows both the primary and secondary model H$\alpha$ line. 

Both radial velocity amplitudes are in agreement with the measurements in \citet{Parsons11}, but have uncertainties reduced by a factor of 3 and 2 respectively.

\subsection{Light curve analysis}
In order to determine the eclipse times of all primary eclipses we created a model that reproduced the shape of the eclipses and then varied the mid-eclipse time and an overall scaling factor to minimise the value of $\chi^2$. All our primary mid-eclipse times are listed in Table~\ref{tab:times}.

We modelled the white dwarfs as spheres using a program that subdivides the visible face of each into 200 concentric annuli. Each annulus contributes an amount to the total stellar flux, which depends upon the limb darkening. The next section gives more details on how we included limb darkening. We accounted for Doppler beaming from the white dwarfs by following \citet{Bloemen11}, and modified the flux by a factor
\begin{equation}
1 - B \, \frac{v_r}{c},
\end{equation}
with $B$ the spectrum-dependent beaming factor and $v_r$ the radial velocity of the star. \textbf{To calculate the beaming factors, w}e used white dwarf model spectra \citep{Koester10} with log($g$) = 7.25 for both white dwarfs, and T = 24000K and T = 12000K for the primary and secondary respectively. These values were chosen after an initial fit as described in this section. With these models, and following \citet{Bloemen11}, the bandpass-integrated beaming factors for the (\uprime, \gprime, \rprime) filters are (1.9, 2.2, 1.3) for the primary white dwarf and (3.4, 3.5, 1.8) for the secondary white dwarf. We did not include gravitational lensing in our models. Due to the similarity of the white dwarfs' masses and radii, and their relatively small separation, the lensing amplification factor near both primary and secondary eclipses is $\sim$ 1.00003, making this effect negligible \citep[for the \textbf{relevant} equations, see][]{Marsh01}.

\begin{table}
\begin{center}
\caption{Properties of CSS\,41177. The J2000 coordinates and the magnitudes are taken from SDSS III DR9, where the magnitudes are the photometric PSF magnitudes.}
\label{tab:properties}
\begin{tabular}{l l}
\hline \hline
property                  & value                 \\
\hline \hline
R.A.                      & 10:05:59.1            \\
Dec.                      & +22:49:32.26          \\
m$_\mathrm{u}$            & 17.32 $\pm$ 0.02      \\
m$_\mathrm{g}$            & 17.29 $\pm$ 0.02      \\
m$_\mathrm{r}$            & 17.62 $\pm$ 0.02      \\
\hline \hline
\end{tabular}
\end{center}
\end{table}

We then normalised each observed light curve individually to reduce any night-to-night variations and used the SDSS magnitudes for CSS\,41177 (see Table~\ref{tab:properties}) to determine the binary's overall out-of-eclipse flux level. We allowed for an additional shift $\delta$, of the secondary eclipse, on top of the 0.5 phase difference with respect to the primary eclipse, by adjusting the phase $\phi$ according to
\begin{equation}
\phi = \phi + \frac{\delta}{2P} (\cos(2 \pi \phi)-1),
\end{equation}
where $P$ is the orbital period. This shift near the secondary eclipse allows for possible R\o mer delays\footnote{Named after the Danish astronomer O. R\o mer, who was the first to realise that the speed of light is finite by observing deviations from strict periodicity for eclipses of Io, a satellite of Jupiter \citep{Sterken05}.} \citep{Kaplan10} and/or a small eccentricity of the orbit.

Our program fits all individual eclipses simultaneously using a Markov chain Monte Carlo (MCMC) method to explore the ten-dimensional parameter space, favouring regions with small $\chi^2$ values of the model with respect to the data. The free parameters in the model were the two radii scaled by the binary separation, $R_1/a$ and $R_2/a$, the white dwarf temperatures, $T_1$ and $T_2$, the radial velocity amplitudes, $K_1$ and $K_2$, the inclination of the binary $i$, the time delay $\delta$, the orbital period $P$ and the zero point of the ephemeris, $T_0$, which was chosen to minimise the correlation with the orbital period. 

Because we have accurate measurements for the radial velocity amplitudes from the analysis of the spectroscopic data we used a prior to constrain them while modelling the photometric data. Given the ten free parameters and combining them with 
\begin{equation}
K_1 + K_2 = \frac{2 \pi a}{P} \; \mathrm{sin}(i),
\end{equation}
and Kepler's equation given by
\begin{equation}
\frac{G (M_1 + M_2)}{a^3} = \frac{4 \pi^2}{P^2},
\end{equation}
the binary's orbital separation $a$ and the white dwarf masses $M_1$ and $M_2$ could also be calculated. Note that $q~=~K_1/K_2~=~M_2/M_1$ and that the surface gravities follow from $g_{1,2}~=~GM_{1,2}/R_{1,2}^2$.

\begin{figure}
\begin{center}
\includegraphics[width=0.48\textwidth]{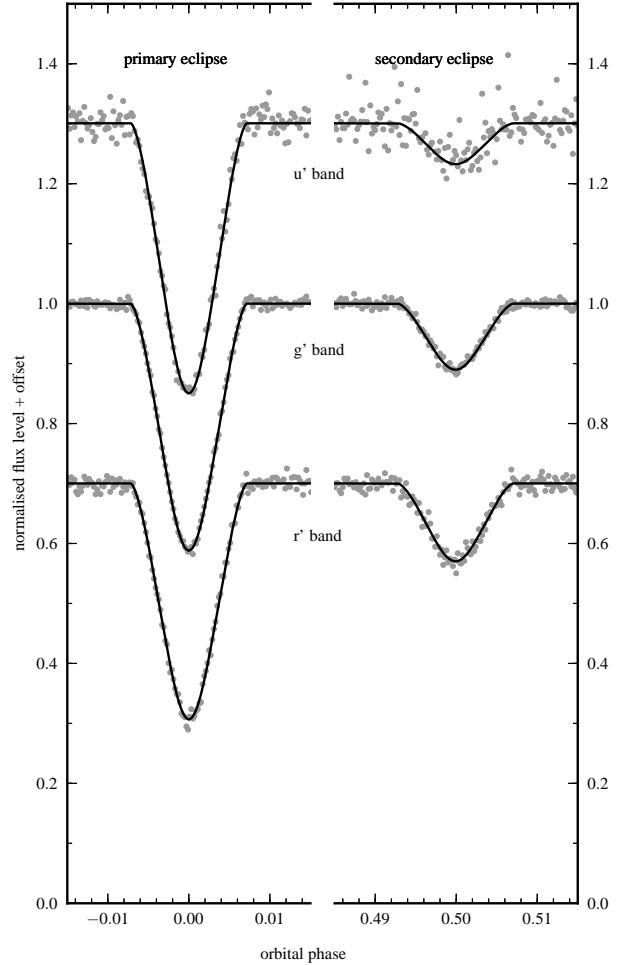}
\caption{ULTRACAM \uprime~(top), \gprime~(middle) and \rprime~(bottom) data, folded with the binary's orbital period, normalised and binned using a binwidth of 0.0002. The \uprime~and \rprime~data are offset by +0.3 and -0.3 respectively, and the \gprime~data are not offset. The black lines show the best model. \emph{Left hand side:} a total of twelve observed eclipses of the primary, hotter white dwarf. \emph{Right hand side:} nine eclipses of the secondary, cooler white dwarf.}
\label{fig:ULTRACAM_eclipses}
\end{center}
\end{figure}

\begin{table*}
\begin{center}
\caption{Mid-eclipse times for the primary eclipses of CSS\,41177. All ULTRACAM times shown are the weighted averages of the \uprime, \gprime~and \rprime~mid-eclipse times.}
\label{tab:times}
\begin{tabular}{l l l l l}
\hline \hline
cycle  & mid-eclipse time  & sampling time  & telescope / instrument & observing conditions \\
number & BMJD(TDB)         & (seconds)      &             & (cloud coverage, seeing) \\
\hline \hline
-3017 & 55599.087790(13)  & 13      & LT / RISE       & clear, seeing 1.8\arcsec \\
-2846 & 55618.926465(15)  & 12      & LT / RISE       & clear, seeing 2.5\arcsec \\
-2544 & 55653.963110(18)  & 12      & LT / RISE       & clear, seeing 3\arcsec \\
-2130 & 55701.9935029(77) & 2 - 8.2 & NTT / ULTRACAM & thin clouds, seeing 1 - 2\arcsec \\
-309  & 55913.257605(22)  & 10      & LT / RISE      & clear, seeing 1.5\arcsec \\
-44   & 55944.0016959(49) & 3 - 6   & WHT / ULTRACAM & thin clouds, seeing 1 - 3\arcsec \\
-43   & 55944.1177028(65) & 3 - 6   & WHT / ULTRACAM & thin clouds, seeing 1.5 - 3.5\arcsec \\
-42   & 55944.233727(10)  & 3 - 6   & WHT / ULTRACAM & clouds, seeing 3 - 10\arcsec \\
-34   & 55945.1618417(55) & 1.5 - 3 & WHT / ULTRACAM & clear, seeing 2\arcsec \\
-33   & 55945.2778585(53) & 1.5 - 3 & WHT / ULTRACAM & some small clouds, seeing 2 - 7\arcsec \\
-16   & 55947.2501251(62) & 1.5 - 3 & WHT / ULTRACAM & thin clouds, seeing 2 - 4\arcsec \\
-8    & 55948.1782455(38) & 2 - 4   & WHT / ULTRACAM & clear, seeing 1 - 2\arcsec \\
-7    & 55948.2942481(58) & 2 - 4   & WHT / ULTRACAM & cloud during egress, seeing 1-5\arcsec \\
-1    & 55948.9903480(47) & 2 - 4   & WHT / ULTRACAM & clear, seeing 1.2\arcsec \\
0     & 55949.1063669(37) & 2 - 4   & WHT / ULTRACAM & clear, seeing 1\arcsec \\
1     & 55949.2223834(39) & 2 - 4   & WHT / ULTRACAM & clear, seeing 1\arcsec \\
3077  & 56306.085857(15)  & 10      & LT / RISE      & clear, seeing 2\arcsec \\
3259  & 56327.200698(16)  & 10      & LT / RISE      & clear, seeing 2\arcsec \\
3568  & 56363.049447(13)  & 10      & LT / RISE      & thin clouds during ingress, seeing 2\arcsec \\
\hline \hline
\end{tabular}
\end{center}
\end{table*}

As a reference our best model is shown in Fig.~\ref{fig:ULTRACAM_eclipses}, on top of the phase-folded and binned data. This model has the lowest chi-squared value, corresponding to a reduced chi-squared of $\chi_{\nu}^2$ = 1.03 (43351 data points, 10 free parameters).

Our final ephemeris is given by
\begin{equation}
\mathrm{BMJD(TDB)} = 55936.3446719(6) + 0.1160154352(15) E,
\label{eq:ephemeris}
\end{equation}
with $E$ the cycle number. The parameters are summarised in Table~\ref{tab:parameters} \textbf{(columns 2 and 3)}, which lists both the mean and rms for each parameter.

\begin{table*}
\begin{center}
\caption{Parameters of the binary and both white dwarfs, \textbf{where the n}umbers in parentheses indicate the uncertainty in the last digit(s). \textbf{The second and third column list the results from the analysis of the spectroscopic and photometric data (see Sect.~\ref{sect:analysis&results}). The last two columns shows the results from two more MCMC analyses, where the limb darkening coefficients (ldc) have been multiplied by 1.05 (fourth column) and where $T_1$ has been fixed (fifth column). }}
\label{tab:parameters}
\begin{tabular}{l l l l l}
\hline \hline
parameter                 & spectroscopy & MCMC analysis    & MCMC with ldc*1.05 & MCMC with fixed $T_1$  \\
\hline \hline
$T_0$ (BMJD(TDB))         & -            & 55936.3446719(6) & 55936.3446719(6) & 55936.3446720(6) \\
$P_{\mathrm{orb}}$ (days) & -            & 0.1160154352(15) & 0.1160154352(15) & 0.1160154351(15) \\
$a$ (\Rsun)               & -            & 0.886(14)        & 0.886(14)        & 0.888(14)        \\
$i$ (deg)                 & -            & 88.97(2)         & 88.96(2)         & 88.95(2)         \\
$\delta$ (seconds)        & -            & -0.79(24)        & -0.80(25)        & -0.78(25)        \\
$M_1$ (\Msun)             & -            & 0.378(23)        & 0.378(23)        & 0.381(23)        \\
$M_2$ (\Msun)             & -            & 0.316(11)        & 0.316(11)        & 0.317(11)        \\
$R_1$ (\Rsun)             & -            & 0.02224(41)      & 0.02227(41)      & 0.02220(41)      \\
$R_2$ (\Rsun)             & -            & 0.02066(42)      & 0.02066(42)      & 0.02087(42)      \\
$T_1$ (K)                 & -            & 24407(654)       & 24362(652)       & 21100            \\
$T_2$ (K)                 & -            & 11678(313)       & 11664(311)       & 10436(21)        \\
log($g_1$)                & -            & 7.321(15)        & 7.319(15)        & 7.325(15)        \\
log($g_2$)                & -            & 7.307(11)        & 7.307(11)        & 7.300(11)        \\
$K_1$ (km s$^{-1}$)       & -176.1(1.1)  & -                & -                & -                \\
$K_2$ (km s$^{-1}$)       & 210.4(6.1)   & -                & -                & -                \\
$\gamma$ (km s$^{-1}$)    & 130.5(0.7)   & -                & -                & -                \\
minimum $\chi^2$          & -            & 44457            & 44457            & 44482            \\
\hline \hline
\end{tabular}
\end{center}
\end{table*}

\subsection{Limb darkening} \label{sect:limb_darkening}
To account for limb darkening of the white dwarfs we used the limb darkening law as first described by \cite{Claret00}, in which the specific intensity across the stellar disc can be calculated using
\begin{equation}
\frac{I(\mu)}{I(1)} = 1 - c_1( 1-\mu^{1/2} ) - c_2( 1-\mu ) - c_3( 1-\mu^{3/2} ) - c_4( 1-\mu^2 ),
\label{eq:claret}
\end{equation}
where $c_1$ - $c_4$ are the limb darkening coefficients, and $\mu$ is the cosine of the angle between the line of sight and the surface normal of the white dwarf, so that $I(1)$ is the specific intensity at the centre of the white dwarf's disc. 

The limb darkening coefficients were recently calculated for a wide range of white dwarf temperatures and surface gravities, for both the Johnson-Kron-Cousins \emph{UBVRI} system and the \emph{ugrizy} filters to be used by the LSST \citep{Gianninas13}. The filter profiles of the ULTRACAM \uprime, \gprime~and \rprime~filters are similar to those of the LSST and therefore we used the coefficients calculated for these LSST filters. We also computed the central specific intensities for these three filters. 

For a given temperature and surface gravity we used a bilinear interpolation between the closest values to calculate all four coefficients and the central specific intensity. These then allowed us to determine the total specific intensity of the white dwarfs, depending on where they are in their orbit and on the fraction of each annulus visible. The total specific intensity\footnote{In units of erg/s/cm$^2$/Hz/sr.} is related to the binary's flux\footnote{In units of erg/s/cm$^2$/Hz.} by a constant, $\alpha$, which we calculated during the MCMC by minimising the difference between the flux defined by the SDSS magnitudes for CSS\,41177 and the product of the total specific intensity with $\alpha$. The constant is related to the solid angle the binary subtends as
\begin{equation}
\alpha = a^2 / D^2,
\end{equation}
where $a$ is the binary's orbital separation and $D$ is the distance to the binary, allowing us to effectively measure the distance from our light curves. For our three filters the resulting distances are $D_u = 481 \pm 37$ pc, $D_g = 473 \pm 35$ pc, and $D_r = 464 \pm 34$ pc. These give an inverse variance weighted value of $D = 473 \pm 35$ pc, with the quoted error similar to those for the individual distances. We believe these errors to be dominated by the uncertainties in the SDSS magnitudes, which is why a weighted error would overestimate the precision with which we can determine the distance. These values are significantly higher than the distance quoted in \citet{Parsons11}, a natural result due to the fact that we also obtained higher temperatures for both white dwarfs.

\textbf{The calculations in \citet{Gianninas13} are based on 1D white dwarf models. \citet{Tremblay13a} have recently shown that the standard 1D mixing-length theory overpredicts surface gravities and, to a lesser extent, temperatures, especially near the values we found for the cooler white dwarf. To assess the effect of using 1D models we compared a 1D and an averaged 3D intensity profile, using a temperature and surface gravity representative of the cooler white dwarf. Full 3D and averaged 3D spectral synthesis produce very similar results at all wavelengths \citep{Tremblay11}, and so the averaged 3D approximation, where the average is performed over constant Rosseland optical depth, is likely to be appropriate for the present study. We found the difference between the two profiles is a factor of 1.05 in the limb darkening coefficients. Running a separate MCMC analysis in which we multiplied each limb darkening coefficient by 1.05, showed that the effect on our values for the parameters is much less than the statistical uncertainty in the parameters (Table~\ref{tab:parameters}, column~4). }

\section{Discussion}
\subsection{Details of the binary orbit} \label{sect:orbit}
For an eclipsing binary on a circular orbit it is often assumed that the primary and secondary eclipses occur exactly half an orbital cycle after one another. However, this is not the case if the two binary components are of unequal mass. The changing distance to the stars at times of eclipses and the finite speed of light cause a shift in the phase of the secondary eclipse. While modelling our light curve we allowed for such a time shift, and found an indication of a small displacement of $\delta$ = -0.79 $\pm$ 0.24 seconds. For CSS\,41177 we theoretically expect the R\o mer delay to be $\delta_R = P(K_2 - K_1) / \pi c$ = 0.36 $\pm$ 0.08 seconds \citep{Kaplan10}, so that the secondary eclipse occurs slightly after phase 0.5. The fact that we measure a delay with the opposite sign to that expected, and that our measurement is 4.6$\sigma$ away from the theoretical prediction indicates that there may be another process at work. 

The measured time delay could be the result of a small eccentricity of the binary's orbit, in which case we can use the measured time delay to constrain the eccentricity. With $\delta_e = 2Pe\cos(\omega)/\pi$ \citep{Kaplan10, Winn10}, where $\omega$ is the argument of pericenter, we obtain an eccentricity of $e\cos(\omega) = -(1.24 \pm 0.38) \times 10^{-4}$, which is a lower limit on the eccentricity. Although not extremely significant, it is certainly possible that the binary did not emerge from the last common envelope phase on a completely circular orbit, or that small perturbations have been induced into the binary's orbit by a third body.

We tried to confirm the measurement of a small eccentricity by fitting the primary white dwarf's radial velocity curve. The result is consistent with an eccentricity of zero, with a 3$\sigma$ upper limit at 0.034. However, this limit is weak compared to the value of $e\cos(\omega)$, which is more than 100x smaller. 

\textbf{The only other precise eccentricity measurement in an eclipsing double white dwarf binary (NLTT~11748) is consistent with a circular orbit, and the measured R\o mer delay for this system agrees with the expected value \citep{Kaplan13}. }

If the \textbf{CSS\,41177} binary orbit is indeed eccentric, apsidal precession will occur. Tidal deformation and rotational distortions of both stars and relativistic processes all contribute to the apsidal precession \citep{Sterne39, Valsecchi12}. The relativistic apsidal precession amounts to 5.6 deg/yr, compared to which the precession rates due to tides and rotation are negligible.

Due to gravitational wave emission the white dwarfs will eventually merge. We calculate the merger time to be $\tau_{\mathrm{m}}$ = 1.14 $\pm$ 0.07 Gyr \citep[][Sect.~4.3]{Marsh95}.

\subsection{Masses, radii, and hydrogen envelopes}
\begin{figure}
\begin{center}
\includegraphics[width=0.49\textwidth]{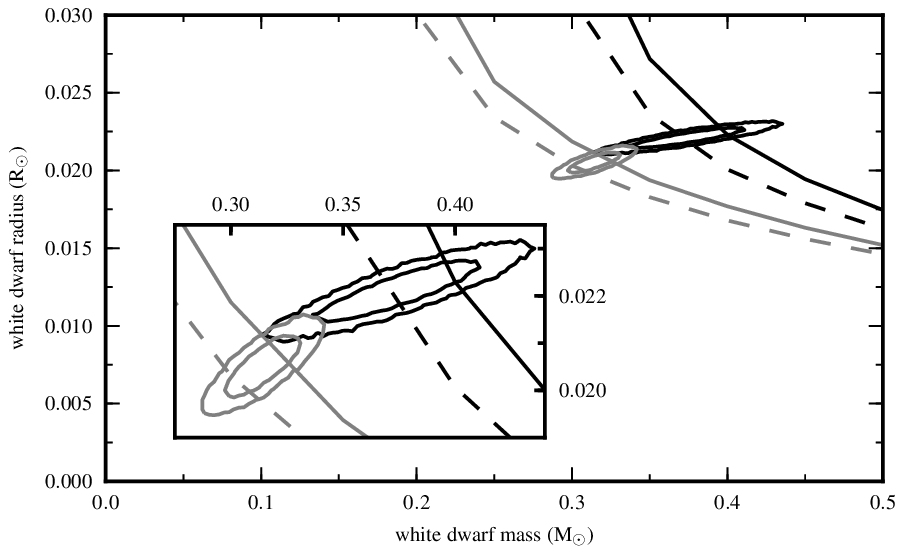}
\caption{The 68 and 95 percentile contours for the masses and radii of the hotter white dwarf (black) and cooler white dwarf (grey) in CSS\,41177. The lines indicate mass-radius relations for hydrogen envelopes of M$_\mathrm{H}$/M$_*$ = 10$^{-4}$ (solid) M$_\mathrm{H}$/M$_*$ = 10$^{-8}$ (dashed), for temperatures of T$_{\mathrm{WD}}$ = 24500 K (black) and T$_{\mathrm{WD}}$ = 11500 K (grey), and all with a metallicity of Z=0.001 \citep{Benvenuto98}. The inset shows an enlargement of the contours and mass-radius relations.}
\label{fig:MvsR}
\end{center}
\end{figure}

\begin{figure*}
\begin{center}
\includegraphics[]{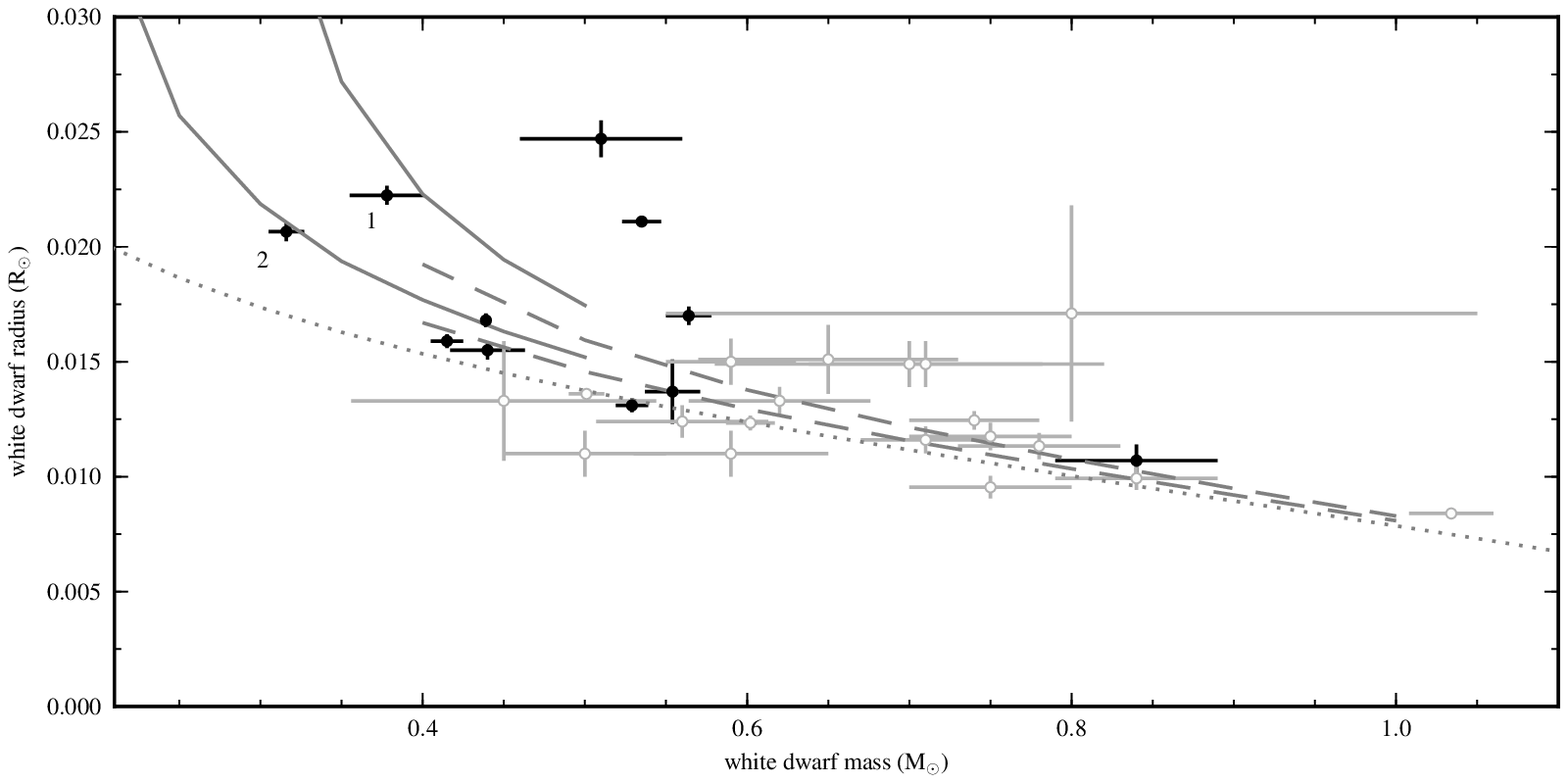}
\caption{Mass-radius diagram showing all white dwarfs with model-independent mass and radius determinations. The hotter and cooler white dwarf in CSS\,41177 are indicated with a 1 and 2 respectively. Other measurements from eclipsing white dwarf binaries are shown as black dots \citep{OBrien01, Maxted04, Maxted07, Parsons10, Parsons12_420, Parsons12_426, Parsons12_419, Pyrzas12}. The grey dots are for white dwarfs based on parallax measurements \citep{Shipman97, Holberg98, Provencal02, Holberg12} and common proper motion systems \citep{Provencal98, Casewell09}. The dotted line shows the zero-temperature mass-radius relations from \citet{VR88}. The solid grey lines are mass-radius relations for helium core white dwarfs with a hydrogen envelope mass of M$_\mathrm{H}$/M$_*$ = 10$^{-4}$, metallicity of Z=0.001, and with T=24500K (upper curve) and T=11500K (lower curve) \citep{Benvenuto98}. The dashed grey lines are for carbon/oxygen core white dwarfs, again with M$_\mathrm{H}$/M$_*$ = 10$^{-4}$ and T=24500K (upper curve) and T=11500K (lower curve) \citep{Wood95}.}
\label{fig:MRdiagram}
\end{center}
\end{figure*}

Typical white dwarf surface gravities are high enough to force elements heavier than hydrogen and helium to settle out of the photosphere on time scales much shorter than evolutionary time scales \citep{Paquette86, Koester09}. As a result, all heavy elements sink below the white dwarf's photosphere, leaving the light elements to form the outer layers. The two low-mass helium white dwarfs in CSS\,41177 have hydrogen envelopes, and are therefore classified as DA white dwarfs. 

Fig.~\ref{fig:MvsR} shows the results from our MCMC analysis for the masses and radii of both white dwarfs, as 68 and 95 percentile joint confidence regions. Also shown in Fig.~\ref{fig:MvsR} are mass-radius relations for hydrogen envelope masses of 10$^{-4}$ (solid lines) and 10$^{-8}$ (dashed lines) of the stellar mass, M$_*$, for both white dwarf temperatures \citep{Benvenuto98}. Our results are in good agreement with both relations, but may favour a low-mass hydrogen envelope for the higher-mass and hotter white dwarf. Observational studies of pulsating white dwarfs suggest that the hydrogen content can be several orders of magnitude smaller than the standard prediction of stellar evolution of 10$^{-4}$ M$_*$, see \citet[][Table~1]{Bradley01}. However, note that those listed white dwarfs are all of significantly higher mass than the CSS\,41177 white dwarfs. 

\begin{figure*}
\begin{center}
\includegraphics[]{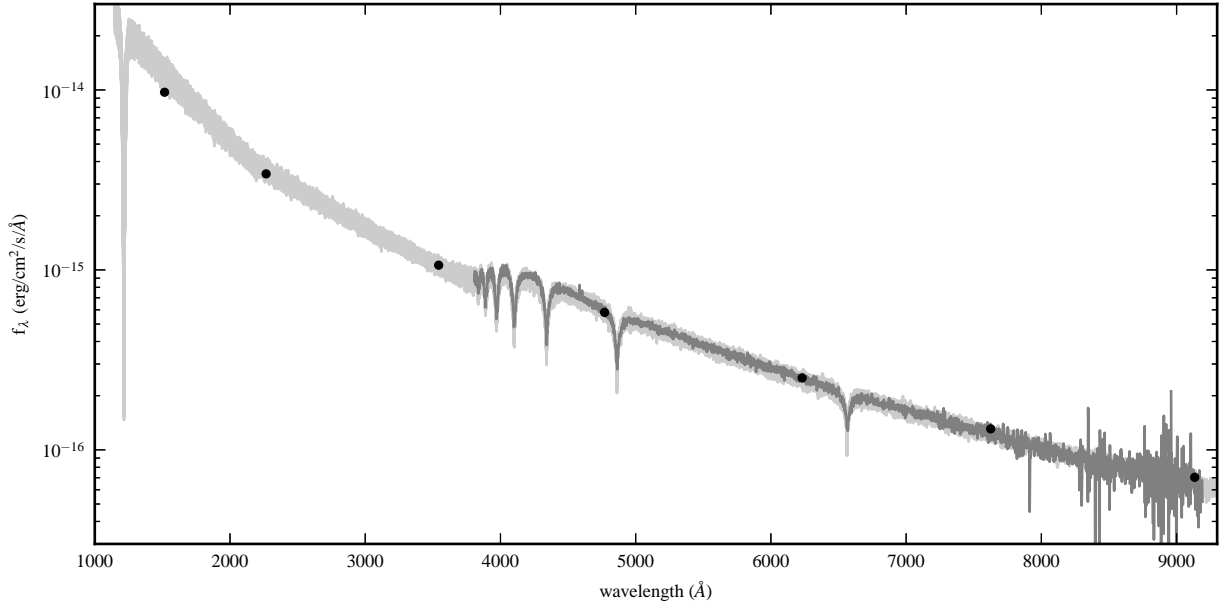}
\caption{Comparison of observed and model spectra of CSS\,41177. The broad light grey line shows 100 different overlapping models, randomly chosen from our MCMC analysis. Each consists of a combination of two white dwarf model spectra corresponding to the MCMC parameters. The total model is scaled to the dereddened \gprime~band SDSS magnitude (m$_g$ = 17.14) and an interstellar extinction of E(B-V) = 0.0339 \citep{Schlegel98} is taken into account using the expressions presented in \citet{Seaton79} and \citet{Howarth83}. The dark grey line shows the observed SDSS spectrum, and the black dots indicate the GALEX far-UV, GALEX near-UV and SDSS \uprime, \gprime, \rprime, \iprime~and \zprime~fluxes (from left to right; errorbars too small to be seen).}
\label{fig:spectrum}
\end{center}
\end{figure*}

The current state of affairs is displayed in Fig.~\ref{fig:MRdiagram}, which shows all highly accurate white dwarf masses and radii, determined independently of mass-radius relations. The two CSS\,41177 white dwarfs (numbered 1 and 2) fall in an area of the mass-radius diagram that has hardly been explored so far, and supply new tests for the theoretical mass-radius relations at low white dwarf masses. In general the measurements agree reasonably well with the models. Note that the solid and dashed grey mass-radius relations are for temperatures of \Teff = 24500 K and 11500 K (upper and lower curves), and that the two notable outliers are both white dwarfs that are significantly hotter (CSS\,03170: \citealt{Parsons12_419} and NN\,Ser: \citealt{Parsons10}).

Using the same models by \citet{Benvenuto98}, we determined the cooling age for the hotter white dwarf to be $\sim$ 50~Myr. The cooler white dwarf is substantially older at $\sim$ 330~Myr. 

\textbf{In general, there is good agreement that double white dwarf binaries like CSS\,41177 go through two phases of mass transfer during their evolution, the second of which is thought to result in a common envelope phase. The nature of the first is less certain, but was most likely a common envelope or stable Algol-like mass transfer \citep{Iben97}. Different binary population synthesis codes agree that both of these evolutionary paths could produce the final CSS\,41177 parameters \citep[][Fig.~A.22 and A.24]{Toonen13}. Under conservative mass transfer, Algol evolution may lead to too small a final orbital separation. Therefore stable non-conservative mass transfer \citep{Woods12}, or a common envelope following the $\gamma$-prescription \citep{Nelemans00} could be a more accurate description of the first phase of mass transfer. }

\subsection{White dwarf pulsations}
In the method explained in Sect.~\ref{sect:analysis&results} we effectively measured the white dwarf temperatures from the depths of the eclipses in the different bands and the temperature-dependent specific intensities from \citet{Gianninas13}. With this approach the temperatures determined are independent of both the SDSS spectrum and model spectra, which \textbf{formed the basis of the temperatures derived} by \citet{Parsons11}. In contrast to their results, we obtain somewhat higher values for both white dwarf temperatures. This is, at least in part, due to the fact that they did not include the effect of reddening. The SDSS spectrum and a group of representative models from our MCMC analysis are plotted in Fig.~\ref{fig:spectrum}\textbf{, with reddening accounted for. The model spectra show that we overestimate the GALEX far-UV flux, indicating that our white dwarf temperatures may be too high. To assess how this influences our conclusions on the masses and radii we ran an additional MCMC analysis in which we fixed the temperature of the hotter white dwarf to $T_1$ = 21100~K, the value found in \citet{Parsons11}. The results are shown in the last column of Table~\ref{tab:parameters}. As expected, the temperature of the cooler white dwarf is also reduced, but the effect on other parameters is well within the uncertainties. }

\textbf{The results from our light curve analysis place} the cooler, secondary white dwarf very close to the \textbf{blue edge of the} DA white dwarf instability strip\textbf{, and the results from our analysis with the fixed low $T_1$ places it near the red edge of the strip}, so we looked for signs of pulsations in the light curve. We inspected the January 2012 data, excluding the single primary eclipse observation from May 2011 to avoid artificial low-frequency signals. Looking at the out-of-eclipse data, we did not find any pulsations with an amplitude exceeding 3.0, 1.0 and 1.1 mmag in the \uprime, \gprime~and \rprime~band. However, the secondary white dwarf's contribution to the flux is strongly diluted by the presence of the primary white dwarf and the flux ratios differ in the three bands. For the \uprime, \gprime~and \rprime~band the primary to secondary flux ratios are 7.7, 4.3 and 3.5 respectively. Correcting for the flux dilution this translates to a non-detection of pulsations with an amplitude exceeding 26.1, 5.3 and 5.0 mmag in the \uprime, \gprime~and \rprime~band. 

The contribution of the secondary to the total amount of flux is highest in the \rprime~band, but white dwarf pulsation amplitudes for non-radial modes increase towards bluer wavelengths. For the $l$ = 1 and $l$ = 2 modes the amplitude in the \gprime~band is $\sim$ 1.4 times higher than the amplitude in the \rprime~band \citep{Robinson95}. Therefore our strongest constraint comes from the \gprime~band. 

\begin{figure}
\begin{center}
\includegraphics[width=0.49\textwidth]{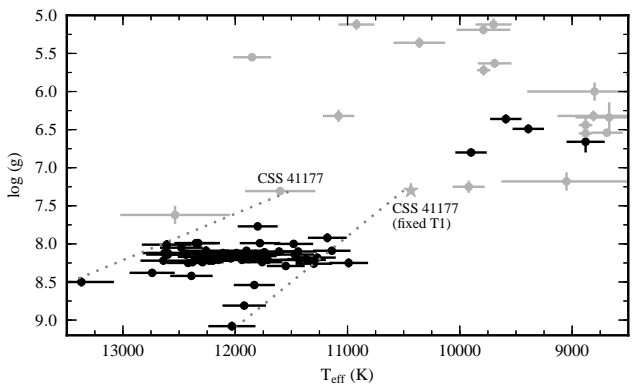}
\caption{ZZ Ceti diagram, showing the position of the cooler secondary white dwarf in CSS\,41177\textbf{, both from our standard analysis and from our analysis with a the fixed low value for $T_1$ = 21100~K}. In the bottom left corner are the pulsating white dwarfs found by \citet{Gianninas11}, and their empirical boundaries for the instability strip. Also in that corner is the most massive pulsating white dwarf found so far, by \citet{Hermes13b}. Towards the top right corner are \textbf{five} pulsating white dwarfs found by \citet{Hermes12} and \citet{Hermes13c} \textbf{(one is off the plot at the right)}. The grey dots represent those white dwarfs that have been found not to be pulsating \citep{Steinfadt12, Hermes12, Hermes13a, Hermes13c}.}
\label{fig:logg_vs_teff}
\end{center}
\end{figure}

Fig.~\ref{fig:logg_vs_teff} shows the log($g$) versus \Teff~diagram, with the white dwarfs found to be pulsating \citep[black dots:][]{Gianninas11, Hermes12, Hermes13c}, and those without pulsations down to 10 mmag \citep[grey dots:][]{Steinfadt12, Hermes12, Hermes13a, Hermes13c}. The latter group now includes the secondary white dwarf of CSS\,41177, \textbf{which is} pushing on the boundary of the instability strip. \textbf{Results from the analysis described in Sect.~\ref{sect:analysis&results} put it at the blue edge of the strip, whereas the results from our analysis with a fixed low value for $T_1$ put the secondary at the red edge. Because \citet{Parsons11} did not account for reddening, the temperatures they derived are an underestimate and the secondary will be somewhat hotter than shown with the star in Fig.~\ref{fig:logg_vs_teff}, placing it even inside the current empirical instability strip. }

Note that the data we have \textbf{are} not ideally suited to search for pulsations, due to the brief sections of out-of-eclipse data. With continuous observations spanning several orbital periods one could push the limit further down, or \textbf{possibly} detect small amplitude pulsations.

\subsection{Orbital period variations}
\begin{figure}
\begin{center}
\includegraphics[width=0.48\textwidth]{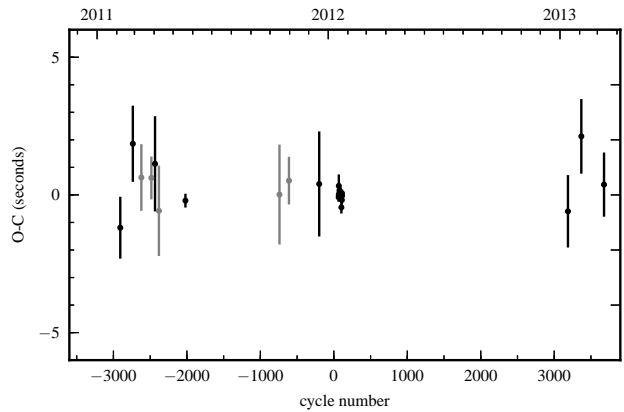}
\caption{Observed minus calculated (O-C) diagram for the primary eclipses of CSS\,41177, including our ULTRACAM and LT/RISE data and 5 eclipse observations from \citet[][shown in grey]{Backhaus12}. We used the ephemeris from Eq.~\ref{eq:ephemeris}. The ULTRACAM eclipses were observed in three filters simultaneously, and the times shown here are the weighted averages of these. See Table~\ref{tab:times} for a list of our eclipse times.}
\label{fig:O-Cdiagram}
\end{center}
\end{figure}

Fig.~\ref{fig:O-Cdiagram} shows the mid-eclipse times of the primary eclipses in an observed minus calculated (O-C) diagram, where the calculation is based on the linear ephemeris in Eq.~\ref{eq:ephemeris}. All our ULTRACAM and LT/RISE data are included. For the ULTRACAM data we show the weighted mean of the \uprime, \gprime~and \rprime~data. All eclipse times are listed in Table~\ref{tab:times}.

This binary \textbf{is} an ideal target for eclipse timing, a method which can be used to detect the presence of circum-binary planetary-like companions \citep[see for example][]{Marsh13, Beuermann10, Beuermann13, Potter11}. This \textbf{is} because the orbital decay through gravitational wave radiation is \textbf{only} $\sim$ 0.5 second over a baseline of 10 years. Also orbital period variations due to magnetic cycles in the stars \citep[the so-called Applegate's mechanism, see][]{Applegate92} are unlikely, as this mechanism is expected to be extremely weak or non-existent in white dwarfs. \textbf{Even i}f the orbit is indeed eccentric, as discussed in Sect.~\ref{sect:orbit}, the apsidal precession of the orbit is too slow to cause a noticeable deviation in the eclipse times over a realistic observational baseline.

\section{Conclusions} \label{sect:conclusions}
We have presented our high signal to noise observations of CSS\,41177 and the analysis of both the spectroscopic and photometric data. The high spectral and temporal resolution and the ULTRACAM observations in three wavelength bands allowed us to accurately model the binary and both white dwarfs, without the need to use mass-radius relations.

The results place these two white dwarfs in a region of the mass-radius diagram that is as yet unexplored; they are the lowest mass white dwarfs with parameters determined without the need to assume a mass-radius relation. Our results agree with white dwarf models for the corresponding temperatures and with standard hydrogen envelopes. \textbf{There are signs} that the hydrogen layers on both CSS\,41177 white dwarfs are thinner than the commonly assumed M$_\mathrm{H}$ = 10$^{-4}$ M$_*$.

Additionally, the secondary, cooler white dwarf explores a new region of the ZZ Ceti diagram. Although its temperature and surface gravity put it close to the boundary of the ZZ Ceti instability strip, we found no significant pulsations in the light curve to a limit of roughly 0.5\% relative amplitude.

We also found an indication that the secondary eclipse does not occur exactly half an orbital period after the primary eclipse. Although the R\o mer delay predicts such an offset, it also predicts that the secondary eclipse will occur late, whereas we measured it to occur early. Therefore the orbit of the two white dwarfs may be slightly eccentric. To measure this effect more precisely more secondary eclipse observations are needed.

\section*{Acknowledgements}
We thank Detlev Koester for the use of the white dwarf model spectra and the referee for helpful comments and suggestions. 
TRM, CMC and BTG were supported under a grant from the UK Science and Technology Facilities Council (STFC), ST/I001719/1. SGP acknowledges support from the Joint Committee ESO-Government of Chile. VSD, SPL and ULTRACAM are supported by the STFC. The research leading to these results has received funding from the European Research Council under the European Union's Seventh Framework Programme (FP/2007-2013) / ERC Grant Agreement n. 320964 (WDTracer).

The William Herschel Telescope is operated on the island of La Palma by the Isaac Newton Group in the Spanish Observatorio del Roque de los Muchachos of the Instituto de Astrof\'isica de Canarias. The Liverpool Telescope is operated on the island of La Palma by Liverpool John Moores University in the Spanish Observatorio del Roque de los Muchachos of the Instituto de Astrof\'isica de Canarias with financial support from the UK Science and Technology Facilities Council. This research has made use of NASA's Astrophysics Data System.

\bibliographystyle{mn_new}
\bibliography{bibliography}

\label{lastpage}
\end{document}